\documentclass[sigplan,screen,nonacm]{acmart}
\AtBeginDocument{%
    \providecommand\BibTeX{{%
        \normalfont B\kern-0.5em{\scshape i\kern-0.25em b}\kern-0.8em\TeX}}}

\usepackage[small,compact,noindentafter]{titlesec} 

\titlespacing\section{0pt}{2pt plus 2pt minus 2pt}{2pt plus 2pt minus 2pt}
\titlespacing\subsection{0pt}{1pt plus 2pt minus 2pt}{2pt plus 1pt minus 1pt}
\titlespacing{\paragraph}{0pt}{2pt plus 0pt minus 1pt}{1.0ex}

\usepackage[inline]{enumitem}
\setlist{noitemsep,topsep=0pt,parsep=0pt,partopsep=0pt}

\usepackage[subtle]{savetrees}

\usepackage{setspace}
\usepackage[margin=0pt,skip=0pt,font={stretch=0.9}]{caption}

\graphicspath{{./plots/}}

\makeatletter
\def\@copyrightspace{\relax}
\makeatother

\begin{document}


    \title[]{Simplifying heterogeneous migration between x86 and ARM machines}

    \author{Nikolaos Mavrogeorgis}
    \email{nikos.mavrogeorgis@ed.ac.uk}
    \affiliation{\institution{University of Edinburgh}}

    \begin{abstract}
        Heterogeneous computing is the strategy of deploying multiple types of processing elements within a single workflow,
        and allowing each to perform the tasks to which is best suited.
        To fully harness the power of heterogeneity, we want to be able to dynamically schedule portions of the code
        and migrate processes at runtime between the architectures.
        Also, migration includes transforming the execution state of the process,
        which induces a significant overhead that offsets the benefits of
        migrating in the first place.
        The goal of my PhD is to work on techniques that allow applications to run on heterogeneous cores
        under a shared programming model,
        and to tackle the generic problem of creating a uniform address space between processes
        running on these highly diverse systems.
        This would greatly simplify the migration process.
        We will start by examining a common stack layout between x86 and ARM binaries, focusing on these two widely spread
        architectures, and later we will try to generalize to other more diverse execution environments.
        On top of that, the performance and energy efficiency of the above effort compared to current approaches will be benchmarked.
    \end{abstract}


\maketitle
    \pagestyle{plain}


    \section{Introduction}\label{sec:introduction}


    In both industry and academia there is the belief that in order to achieve higher performance, energy efficiency
    and lower cost, attention must be given to specialized hardware~\cite{indatacentertpu}.
    Combining this special-purpose hardware (e.g.~GPUs, TPUs, accelerators) with general-purpose hardware (e.g.~CPUs),
    and reconfigurable integrated circuits (e.g.~FPGAs),
    leads to heterogeneous architectures,
    allowing mixed workloads to be executed in the most suitable hardware~\cite{AMD2011}.
    The different processing units may share the same ISA, like in the big.LITTLE architecture~\cite{jeff2012big},
    or even employ heterogeneous-ISAs.
    A heterogeneous-ISA approach yields even more diversity to exploit, like different code density, register depth and
    width or decode instruction complexity~\cite{venkat2014harnessing}.
    For example, smartNIC and smartSSD processors are increasingly adopted in the data center~\cite{yanfang2017,kang2013},
    and they are usually paired with x86 host processors,
    which could lead to large-scale live migration between architectures.


    Whatever the type of heterogeneity supported, a shared memory programming (SMP) model is usually the most convenient solution
    for the programmer.
    Moreover, bus standards like OpenCAPI~\cite{opencapi} have emerged, aiming to provide improved coherency and synchronization
    in heterogeneous systems, favoring SMP programming models.
    Therefore, we will focus on the setting of SMP models in heterogeneous-ISA environments.

    An approach like Popcorn~\cite{barbalace2017breaking}, allows applications to migrate during runtime at equivalence
    points of computation in the program~\cite{Lyerly2016}, but requires first the stack transformation of the program
    between architectures.
    This creates an overhead to the migration,
    which grows linearly with the number of stack frames, and the number of variables,
    but also poses additional engineering effort for creating the transformation toolchain.
    Also, the above technique generates two fat binaries, with some metadata necessary for migration,
    which are more difficult to deploy in embedded systems.
    In order to harness the power of various heterogeneous techniques, it is necessary to have efficient process migration
    at runtime, otherwise the migration costs will offset the benefits of heterogeneity.


    One effective way to exploit heterogeneity would be to circumvent the stack transformation at runtime by imposing a
    uniform stack layout between the target architectures.
    The focus currently is between the x86-64~\cite{Devices2012} and ARMv8~\cite{Limited2020} target ISAs. 
    The first fundamental part of this idea, is to keep the Application Binary Interface (ABI) the same between the
    architectures.
    On top of that, we need to explore the effect of various optimizations (target-dependent or not) that affect spills and
    refills in memory.
    To this end, we modify\, the LLVM's backend infrastructure~\cite{lattner2008introduction}, controlling the emitted
    assembly instructions, and tune the optimization flags accordingly.


    The first difference from existing works like bit.LITTLE, is that we focus on heterogeneous-ISA architectures,
    instead of homogeneous ones.
    In the context of ISA heterogeneity, some of the current approaches require the transformation at runtime of the
    stack from one architecture to another, in order to migrate a thread, like in the case of
    Popcorn software~\cite{barbalace2015popcorn} or Venkat's work~\cite{bhat2016harnessing}.
    We will try to avoid this transformation.
    Ultimately, our goal would be to extend this work to multiple architectures, instead of relying solely on the
    64-bit versions of x86 and ARM, or on other commonly used combinations, as described in Section~\ref{sec:related-work}.

    \section{Overview of the proposed work}\label{sec:overview}

    Our initial focus is to find ways to overcome the stack transformation overhead at runtime.
    The way an application interacts with its stack, doing load and store instructions, is determined by various factors,
    such as the ABI, the optimizations done at compile-time and the target-dependent code generation.

    The first step, was to modify LLVM's backend for the x86-64 and Aarch64 targets.
    First of all, the number of registers available in the architecture (register depth) affects the register
    pressure, i.e.~the number of simultaneously live variables at an instruction) and, consequently, the frequency of
    spills and refills in the stack~\cite{Braun2009}.
    Therefore, we modified the architecture backend, to get the same register depth.
    We have, also, mapped the various registers one-to-one, taking into consideration their occasional target specific functionality,
    e.g.~the platform register in ARMv8.
    Moreover, the procedure calling convention dictates the number of registers that will be spilled at the boundary
    of the stack frame as arguments.
    In the same way, it enumerates the callee-saved registers, which will also be saved in the stack frame in case
    they are to be modified.
    Hence, we implemented the same calling conventions for both targets.

    However, even if these differences are completely resolved,
    we still have to observe the dynamic behaviour of the executed application, and its interaction with the stack.
    Ideally, we aim to generalize our solution to other architectures, with different register depths, pointer sizes and
    other architecture specifics.

    \section{Preliminary results}\label{sec:preliminary-results}

\begin{figure}[t]
        \includegraphics[width=0.48\textwidth]{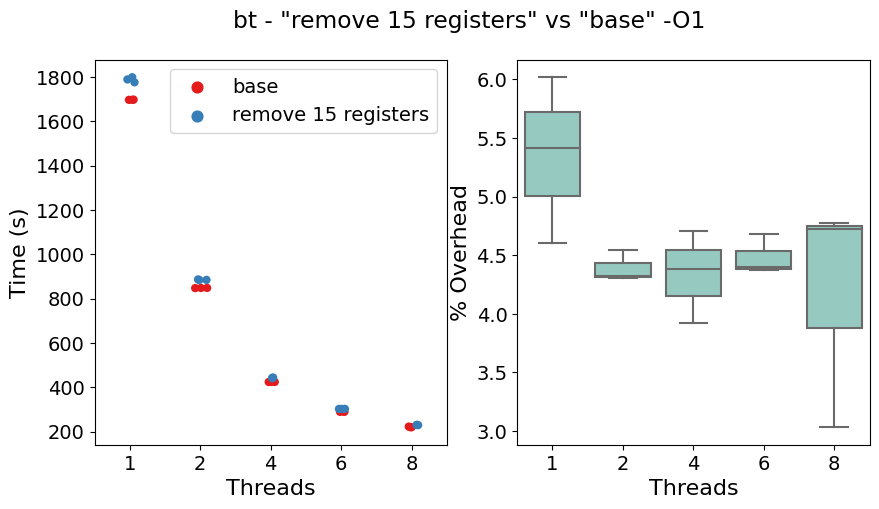}\hfill
    \caption{BT results for a modified AArch64 target}\label{fig:figure}
    \end{figure}

    A main concern is to ensure that forcing some characteristics on an architecture does not impact performance
    a lot during program execution, given that we are missing some architecture-specific optimizations. 
    For instance, we have removed 15 of the general purpose registers (GPRs) of the Aarch64 LLVM target, which had 32 GPRs,
    to match the 17 GPRs of the x86-64 architecture (including the stack pointer).
    So far, we have run experiments only on integer benchmarks, for the C language.
    Also, we focused on compute- and\, memory-intensive benchmarks.
    Running the NAS Parallel Benchmarks \cite{Bailey1991}, we measure the overhead of removing the registers for Aarch64 and
    show some sample plots for one benchmark in Figure~\ref{fig:figure}.
    All in all, we observed no more than 6\% of overhead in the \textit{-O1} flag for the benchmark suite.

    \section{Work to be done}\label{sec:work-to-be-done}


    Now we will describe the necessary changes to the LLVM backends of the targets.
    Where a register in an architecture does not have a reciprocal to the other architecture (e.g.~there is no platform
    register in x86-64), we plan to either emulate the same behaviour, or ignore it if it has no important role \,
    to our prototype.

    In addition, there are some fundamental differences between our targets that lead to stack mismatch during execution.
    For example, the x86 architecture is CISC and includes operations between the memory and the registers.
    On the contrary, the ARM architecture is RISC and employs the load-compute-store philosophy,
    which creates additional register pressure, especially now that we have reduced the number of ARMv8 registers to
    match the number of x86-64 registers.
    Therefore, we need to empirically identify stack layout differences at runtime and pinpoint the causes of these
    differences.

    Following the above, we may have to mitigate those differences, either at the intermediate representation (IR) level (e.g.~with a flag like
    \textit{mem2reg}, to reduce unnecessary loads),  or at a target-dependent level.
    If tuning the flags is not sufficient, we may again need to modify the code generation of the respective
    backend, to emulate the desired behaviour.

    After this work has matured enough, we will try to generalize our approach by analyzing the fundamental properties
    of an architecture that shape its stack at runtime.
    Consequently, we will examine the feasibility of\, an architecture design and/or a specification of a more universal ABI,
    that allows us to expand to more diverse targets, like CPU/DPU systems.
    In order to address architectures with smaller number of registers,
    we may have to emulate some registers by consuming more stack area.

    Finally, we will check experimentally that implementing a similar address space does not lead
    to performance issues, for execution on only one architecture.

    \section{Related Work}\label{sec:related-work}

    Having a unified address space in heterogeneous environments, is not a new idea.
    AMD started originally the Heterogeneous System Architecture (HSA) framework, supporting a
    single address space accessible to both CPU and GPU, through virtual address translation~\cite{Kyriazis2012}.
    Still, it is the programmer's responsibility to identify and schedule the workloads statically.
    In the Popcorn compiler toolchain and state transformation runtime~\cite{barbalace2017breaking},
    a separate binary is created for each architecture,
    which is augmented with metadata by the middle- and back-end LLVM passes, necessary for the state transformation.
    When a migration occurs, the runtime converts all function activations from the source ISA format to the destination
    ISA format~\cite{Lyerly2016}.
    In another case, Venkat\cite{venkat2014harnessing} and DeVuyst~\cite{devuyst} use dynamic binary translation for one fat binary,
    until the binary reaches a migration point and state transformation takes place.
    These last efforts also included smaller register widths, like 32 bits, but we foresee that architectures with at
    least 64 bits will dominate, hence we focus on them for now.

    

    \section{Conclusion}\label{sec:conclusion}
    This paper presents the outline of our proposed approach, for simplifying heterogeneous ISA migration.
    Our plan is to come up with solutions that create a uniform address space between the processes,
    whereas eliminating the overhead of state transformation at runtime.
    Preliminary results indicate that enforcing a similar stack layout,
    through an identical ABI between x86-64 and Aarch64 targets,
    does not deteriorate performance significantly, while it paves the way for actually creating identical
    stack layouts.
    An ultimate goal would be to extend this work, in order to ease portability
    of migration to various targets, other than x86-64 and ARMv8.

\begin{acks}
The work has been supervised by Dr Antonio Barbalace at the University of Edinburgh.
The PhD started in 2020 and thesis submission is expected in 2023.

\end{acks}

\bibliographystyle{ACM-Reference-Format}
\typeout{}
\bibliography{bib}

\end{document}